\documentclass[12pt]{article}
\usepackage[xdvi]{graphicx}
\usepackage{epsfig}
\usepackage{amsfonts,amssymb,amsmath}
\usepackage{subfigure}
\newcommand{\mn}{\ensuremath{{\mu\nu}}}
\newcommand{\al}{\ensuremath{\alpha}}
\newcommand{\ga}{\ensuremath{\gamma}}

\newcommand{\La}{\ensuremath{\Lambda}}

\newcommand{\s}{\ensuremath{\sigma}}

\newcommand{\mc}{\ensuremath{\mathcal}}

\newcommand{\be}{\begin{equation}}
\newcommand{\ee}{\end{equation}}

\newcommand{\bea}{\begin{eqnarray}}
\newcommand{\eea}{\end{eqnarray}}
\newcommand{\ba}{\begin{eqnarray}}
\newcommand{\ea}{\end{eqnarray}}

\begin{document}

\rightline{hep-th/0512083}
\vskip 1cm

\begin{center}
{\Large \bf A covariant approach to  braneworld holography}
\end{center}
\vskip 1cm
\renewcommand{\thefootnote}{\fnsymbol{footnote}}

\centerline{\bf Antonio Padilla\footnote{antonio.padilla@utsa.edu, padilla@ffn.ub.es}}
\vskip .5cm

\centerline{\it Department of Physics and Astronomy}
\centerline{\it University of Texas San Antonio, 6900 N Loop 1604 W, TX 78249, USA}
\centerline{\it and}
\centerline{\it Departament de Fisica Fonamental}
\centerline{\it Universitat de Barcelona, Diagonal 647, 08028 Barcelona, Spain}

\setcounter{footnote}{0} \renewcommand{\thefootnote}{\arabic{footnote}}


\begin{abstract}
Exact holography  for cosmological
branes in an AdS-Schwarzschild bulk was first introduced in
hep-th/0204218. We extend this notion to include all co-dimension one
branes moving in non-trivial bulk spacetimes. We use a covariant
approach, and show  that the bulk Weyl tensor projected on to the
brane can always be traded in for ``holographic'' energy-momentum on
the brane. More precisely, a brane moving in a non-maximally symmetric
bulk has exactly the same geometry as a brane moving in a maximally
symmetric bulk, so long as we include the holographic
fields  on the brane. This correspondence is exact in that it works
to all orders in the brane energy-momentum tensor.
\end{abstract}

\newpage
\section{Introduction}
Inspired by the entropy formula of black holes,
the
holographic principle asserts that there is a duality between gravity
in $n$ dimensions and a gauge theory in $n-1$ dimensions. The first
concrete example of this was Maldacena's AdS/CFT
correspondence, in which IIB supergravity on
$\textrm{AdS}_5 \times \textrm{S}^5$ was found to be dual to
$\mathcal{N}=4$ super Yang-Mills theory on the boundary~\cite{adscft, Witten:adscft}.

This remarkable idea can be studied in braneworld theories~\cite{RS2, locally}. In the
single brane Randall-Sundrum model~\cite{RS2}, we can  think of gravity in the
asymptotically anti de Sitter (AdS)  bulk as
being dual to a conformal field theory (CFT) on the brane/boundary \cite{Gubser}. The CFT has a UV cut-off and  is
coupled to gravity. In a very nice paper \cite{Savonije}, Verlinde and Savonije examined
cosmological branes moving in an $n$-dimensional AdS-Schwarzschild bulk. In the limit
that the brane was close to the AdS boundary, they showed that the
brane cosmology agreed with the standard cosmology in $(n-1)$
dimensions. Furthermore, the braneworld observer would see the bulk
black hole {\it holographically} as dark radiation. Some time later, James Gregory and I noted that a holographic description
held even when the brane was deep inside the bulk, far away from the
boundary of AdS~\cite{Gregory:exact}. Briefly speaking, an observer living on an empty
brane in an AdS Schwarzschild bulk experienced {\it exactly the same}
evolution as an observer living on a non-empty brane in a
maximally symmetric AdS bulk. The latter brane is non-empty in the
sense that the fields of a ¨dual¨ gauge theory are sitting on the brane. The
field theory is not conformal in general, although it approaches a CFT
as the brane approaches the AdS boundary. Because of the remarkable
exactness in the correspondence we found, we later dubbed this work
{\it exact braneworld holography}~\cite{Padilla:thesis}.

In this paper, we will extend this notion of exact braneworld
holography to include a much larger class of braneworlds. We will
adopt a covariant approach to show
that the geometry of any  brane in any non-maximally symmetric bulk is the
same as the geometry of a brane in a maximally symmetric bulk,
provided we add some ¨holographic¨ matter to the brane. This
holographic picture could be very useful in that we manage to entirely
avoid the troublesome Weyl term projected on to the brane~\cite{cov}.

The rest of this paper is organised as follows: in the next section we
briefly review exact holography for cosmological branes. In section 3,
we show how exact holography is extended to arbitrary brane and bulk
geometries using a covariant approach and the Brown and York (BY)
stress-energy tensor~\cite{BY}. In section 4, we  discuss some
properties of the holographic energy-momentum tensor, and suggest ways
of calculating it explicitly. Section 5 contains some
concluding remarks, and a discussion of the generalisation of this
work to Lovelock gravities.

\section{Exact holography for cosmological branes}

We start by briefly reviewing precisely what we mean by exact holography for
cosmological branes (for more details,  see~\cite{Gregory:exact, Padilla:thesis}). Consider an $(n-1)$-dimensional brane moving in a
maximally symmetric
$n$-dimensional AdS bulk. In global coordinates the bulk metric is
given by
\be \label{bulk}
ds^2=-V(a) dT^2 + \frac{da^2}{V(a)}+a^2 q_{ij}dx^i dx^j
\ee
where
\be
V(a)=k^2a^2+1
\ee
and $q_{ij}$ is the metric on a unit $(n-2)$-sphere. The brane is the following embedding in the bulk geometry
\be
T=T(t), \qquad a=a(t), \qquad x^i=y^i
\ee
where
\be
-V (a) \left[ \dot T(t)\right]^2+\frac{ \left[\dot a(t)\right]^2}{V(a)}=-1
\ee
This ensures that the brane metric is Friedmann-Robertson-Walker. Since the brane is
cosmological, we assume that its energy-momentum is made up of tension, $\s$, and additional matter with
energy density, $\rho$, and pressure, $p$. The Friedmann equation
is~\cite{Binetruy:unconventional1, Binetruy:unconventional2}
\be \label{eq:frwexact1}
H^2=\left(\frac{\dot a}{a}\right)^2=\frac{2\La_{n-1}}{(n-2)(n-3)}-\frac{1}{a^2}+\frac{16 \pi G_{n-1}}{(n-2)(n-3)}\rho\left[1 +
\frac{\rho}{2\s}\right].
\ee
where $\La_{n-1}$ and $ G_{n-1}$ are the braneworld cosmological
constant and Newton's constant respectively. Note that this takes the form of the $(n-1)$-dimensional standard cosmology when
$\rho \ll \s$.

Now consider a brane with no additional matter, moving in an AdS black
hole bulk. The bulk metric is now given by (\ref{bulk}) with
\be
V(a)=k^2 a^2+1-\frac{\mu}{a^{n-3}}.
\ee
The black hole mass is proportional to $\mu$. We can embed a
cosmological brane in a similar way, and find that in this case the Friedmann equation is given by
\be \label{eq:frwexact2}
H^2=\frac{2\La_{n-1}}{(n-2)(n-3)}-\frac{1}{a^2}+\frac{\mu}{a^{n-1}}.
\ee
In~\cite{Gregory:exact}, we showed how we can calculate {\it exactly}
the energy density of the black hole bulk, $\rho_\textrm{holog}$, measured by an observer on the
brane -- this can be done {\it without} assuming that the brane is
near the AdS boundary. $\rho_\textrm{holog}$ is given in terms of $\mu$,
so we can rewrite the Friedmann equation (\ref{eq:frwexact2}) to give
\be
H^2=\frac{2\La_{n-1}}{(n-2)(n-3)}-\frac{1}{a^2}+\frac{16 \pi G_{n-1}}{(n-2)(n-3)}
 \rho_\textrm{holog}\left[1 + \frac{\rho_\textrm{holog}}{2\s}\right].
\ee
This takes exactly the same form as the Friedmann
  equation~(\ref{eq:frwexact1}) for the brane moving in maximally symmetric AdS space
  with additional matter on the brane. We can therefore think of
  $\rho_\textrm{holog}$ as being the energy density of a field theory
  living on the brane. This field theory is dual to the AdS black hole
  bulk, although it is no longer conformal.  We think of the dual
  field theory on the brane as being cut off in the ultra violet -- this
  cutoff disappears as we go closer and closer to the AdS boundary,
  and we approach a conformal field theory.  In this case, we are not
  assuming that the brane is near the boundary, so the cutoff can be
  significant.
\section{Exact holography for  all co-dimension one branes}

We will now show the result reviewed in the previous section can be
generalised to a much broader class of brane geometries. We make use of the covariant
formalism of~\cite{cov}, and are able to generalise the notion of
exact holography in a remarkably clear and simple way.

Consider an $(n-1)$-dimensional brane moving in an $n$-dimensional
bulk. We will assume for simplicity that we have $\mathbb{Z}_2$ symmetry across
the brane. This means that the brane splits the  bulk into two identical
domains. Each domain can be thought of as a manifold $\mathcal{M}$,
with a boundary $\partial \mathcal{M}$ that coincides with
the brane.

Now for some notation. The bulk metric is given by
\be
ds^2=g_{ab}dx^a dx^b
\ee
As in the previous section, we can think of the brane as an embedding in
the bulk geometry
\be
x^a=X^a(y^\mu).
\ee
We use this to determine the tangents to the brane
\be
  V^a_\mu=
\frac{\partial X^a}{\partial y^\mu}
\ee
The induced metric on the brane is therefore given by
\be
\gamma_{\mu\nu}= g_{ab}V^a_\mu V^b_\nu
\ee
We will also denote the normal to the brane by $n^a$. This enables us
to define the brane extrinsic curvature
\be
K_{\mu\nu}=\nabla_{(a} n_{b)}V^a_\mu V^b_\nu
\ee
We are  now ready to define the
action describing our braneworld scenario
\be \label{action}
S=2S_g +S_m
\ee
where
\begin{equation}
S_g = M^{n-2}\left[\int_\mathcal{M} d^5 x \sqrt{-g}
\left(R(g)-2\Lambda\right) + \int_{\partial \mathcal{M} }d^4
y\sqrt{-\gamma} 2K\right]
\end{equation}
\be
S_m = \int_\textrm{brane} d^4 y\sqrt{-\gamma} L_m
\ee
Here $M$ is the bulk Planck mass, and $\Lambda$ is the bulk
cosmological constant. There is no additional matter in the bulk,
although there is an arbitrary matter distribution on the brane with
Lagrangian $L_m$. Note that we have two copies of $S_g$ in
the action (\ref{action}) because we have two copies of the bulk $\mathcal{M}$.

It is worth pointing out at this point that we have used the so-called
``Trace-K'' form for the gravitational part of the action $S_g$~\cite{GH}. This
is in keeping with our notion of the brane  forming the boundary of
the bulk spacetime. The Gibbons-Hawking term ensures that the correct bulk and brane equations of motion
are obtained from  varying the action
with respect to the bulk and brane metrics respectively. This approach
is in contrast to the more common approach used in the braneworld
literature, where the brane is regarded as a delta-function source in
the Einstein equations. The two approaches are entirely equivalent at
the level of the equations of motion, as of  course they should
be. The distinction lies at the level of the action: the
Gibbons-Hawking term is not required in the more common approach,
whereas it is required in the ``variational'' approach we use here. We
have chosen this ``variational'' approach because it enables us to see
the generalisation of exact holography much more easily.

We now proceed with varying the action with respect to the bulk and
brane metrics~\cite{dynamicdw}. The bulk equations of motion are
just the Einstein equations with a cosmological constant \be
\label{Einstein} \frac{2}{\sqrt{-g}}\frac{\delta S}{\delta g^{ab}}=0
\implies R_{ab}-\frac{1}{2} Rg_{ab}=-\Lambda g_{ab}. \ee whereas the
brane equations of motion are the Israel equations \be
\label{Israel} \frac{2}{\sqrt{-\gamma}}\frac{\delta S}{\delta
\gamma^{\mu\nu}}=0 \implies 4M^{n-2}\left( K_{\mu\nu}-K
\gamma_{\mu\nu} \right)=T^{(m)}_{\mu\nu} \ee where \be
T^{(m)}_{\mu\nu}=-\frac{2}{\sqrt{-\gamma}}\frac{\delta S_m}{\delta
  \gamma^{\mu\nu}}
\ee
Since we are mainly interested in the dynamics felt by an observer living
on the brane, we will make use of the Gauss-Codazzi equations.
\be \label{Gauss}
\mathcal{R}_{\mu\nu\alpha\beta}(\gamma)=R_{abcd}(g)V^a_{\mu}V^b_{\nu}V^c_{\alpha}V^d_{\beta}+K_{\mu\alpha}K_{\nu\beta}-K_{\mu\beta}K_{\nu\alpha}
\ee
\be \label{codazzi}
D^\mu (K_{\mu\nu}-K\gamma_{\mu\nu})=R_{ab}n^aV^b_\nu
\ee
where $D_\mu$ and $\mathcal{R}_{\mu\nu\alpha\beta}(\gamma)$ are the
covariant derivative and Riemann tensor for the brane geometry. Given the Israel equations (\ref{Israel}), and the fact that $R_{ab}
\propto g_{ab}$, we can use the Codazzi equation (\ref{codazzi}) to show that
energy on the brane is conserved
\be
D^\mu T^{(m)}_{\mu\nu}=0
\ee

The Einstein equations (\ref{Einstein}) imply that the bulk Riemann
tensor takes the following form \be \label{riemann}
R_{abcd}(g)=C_{abcd}+\frac{2\Lambda}{(n-1)(n-2)}(g_{ac}g_{bd}-g_{ad}g_{bc})
\ee where $C_{abcd}$ is the bulk Weyl tensor. This vanishes when we
have maximal symmetry.  Inserting (\ref{riemann}) into the Gauss
equation (\ref{Gauss}), contracting, and making use of the Israel
equations (\ref{Israel}), we obtain the following formula for the
brane Ricci tensor~\cite{cov} \be \label{ricci}
\mathcal{R}_{\mu\nu}(\gamma)=-E_{\mu\nu}+\frac{2\Lambda}{n-1}\gamma_{\mu\nu}-\left(\frac{1}{4M^{n-2}}\right)^2\left[T^{(m)}_{\mu\alpha}T^{(m)}{}^\alpha_\nu-\frac{T^{(m)}}{n-2}T^{(m)}_{\mu\nu}\right]
\ee where $E_{\mu\nu}=C_{abcd}V^a_\mu n^b V^c_\nu n^d$ represents a
non-local contribution coming from the bulk Weyl tensor. For the
cosmological brane discussed in the previous section, it corresponds
to the term proportional to $\mu$ in equation
(\ref{eq:frwexact2})~\cite{globalbw}. It is this quantity that we
would like to interpret holographically. Can we reinterpret it as
some holographic fields living on the brane?

As in~\cite{Gregory:exact}, the idea is that we calculate the
energy-momentum of the bulk measured by an observer living on the
brane. How might we go about doing this? It is well known that there
is no local definition for the
stress-energy-momentum (SEM)  of the bulk gravitational field~\cite{BY}. One needs to adopt a ``quasi-local'' definition on the boundary of
a given region. Furthermore, if we wish to derive a global quantity,
such as the total SEM in the bulk, one does so by considering the
limit of the quasi-local SEM measured by observers on the boundary of
the entire bulk. In our case, this boundary corresponds to the brane,
so we immediately arrive at the bulk SEM measured by observers on the brane.

We will  use Brown and York's definition for the  quasi-local
stress-energy tensor, $T_{\mu\nu}^\textrm{BY}$ ~\cite{BY}. We believe this is a
compelling definition since it enables us to associate the following
conserved charge, with a Killing vector, $\xi^\mu$ on the boundary.
\be \label{Q}
Q(\xi)=\int_S d^{n-2}\zeta \sqrt{\lambda}~ u^\mu T_{\mu\nu}^\textrm{BY}
 \xi^\nu
\ee
where $S$ is a spacelike surface lying in $\partial \mc{M}$, with normal
$u^\mu$, and induced metric
\be
\lambda_{ij}=\gamma_{\mu\nu}\frac{\partial y^\mu}{\partial \zeta^i}\frac{\partial y^\nu}{\partial \zeta^j}
\ee
Making use of the Brown and York stress-energy tensor requires us
to define a suitable background spacetime\footnote{From now on,  we  will label all  background quantities
with a ``bar'' , as will become obvious.}. The background we choose
satisfies the following
two properties~\cite{H+H, BY, Gregory:exact, Padilla:thesis}:
\begin{itemize}
\item
the bulk, $\mathcal{\bar M}$,
is maximally symmetric so that
\be \label{bgriemann}
\bar R_{abcd}(\bar g)=\frac{2\Lambda}{(n-1)(n-2)}(\bar g_{ac} \bar
g_{bd}-\bar g_{ad}\bar g_{bc})
\ee
\item
the boundary, or ``cutoff'' surface, $\partial\mathcal{\bar M} $, must
have exactly the same geometry as the brane. In other words
\be
\bar \gamma_{\mu\nu}=\gamma_{\mu\nu}
\ee
\end{itemize}
Some comments are in order here. Firstly, we believe that the choice
of a maximally symmetric background is a natural one, and indeed the
one that is most often used in calculating, say, the mass of a black
hole spacetime. Furthermore, the Weyl term $\bar C_{abcd}=0$, for this
background. Recall that it is the Weyl term
$E_{\mu\nu}$ in (\ref{ricci})  that we are trying to understand
holographically. It is appropriate that we should choose a background
for which this term is absent.

Secondly, we have followed the presciption of~\cite{BY, H+H} in
demanding that the background be cut-off at a surface whose induced
metric is identical to the brane metric. As with the brane, we can think of the cutoff,  $\partial\mathcal{\bar
  M} $, as a surface embedded in the background bulk. In principle it
  is not always possible to find an embedding with precisely the
  desired  geometry. However, we have seen that it is always possible
  for the cosmological branes described in the previous section. We
  shall proceed under the assumption that a suitable cutoff surface
  can indeed be found.

Now that we have defined a background, we can define the {\it
  physical} action for the bulk~\cite{H+H}
\be
S_{phys}=S_g-\bar S_g.
\ee
Here the background action is given by
\be
\bar S_g= M^{n-2}\left[\int_\mathcal{\bar M} d^5 x \sqrt{-\bar g}
\left(\bar R(\bar g)-2\Lambda\right) + \int_{\partial \mathcal{\bar M} }d^4
y\sqrt{-\gamma} 2\bar K\right]
\ee
We are now ready to calculate the BY stress-energy tensor of  the bulk as
  measured by an observer on the brane
\be
T^\textrm{BY}_{\mu\nu}=-\frac{2}{\sqrt{-\gamma}}\frac{\delta S_{phys}}{\delta
  \gamma^{\mu\nu}}=-2M^{n-2}\left( K_{\mu\nu}-K \gamma_{\mu\nu}
  \right)+2M^{n-2}\left(\bar K_{\mu\nu}-\bar K \gamma_{\mu\nu} \right)
\ee
We now associate this with the {\it holographic} energy-momentum
  tensor, $T^{(h)}_{\mu\nu}=T^\textrm{BY}_{\mu\nu}$. Making use  of the Israel equation (\ref{Israel}), we see that
\be
4 M^{n-2}\left(\bar K_{\mu\nu}-\bar K \gamma_{\mu\nu} \right)
=T^{(m)}_{\mu\nu}+2T^{(h)}_{\mu\nu}
\ee
This equation corresponds to the Israel equation for the cutoff
  surface, $\partial\mathcal{\bar
  M} $ moving in the background bulk. Note that it
  is behaving like a brane containing the original matter,
  $T^{(m)}_{\mu\nu}$, plus some additional holographic matter,
  $T^{(h)}_{\mu\nu}$. There are two copies of the holographic matter
 because  there were two copies of $\mathcal{M}$.

Because the background, $\mathcal{\bar M}$, is maximally symmetric, there
  is no bulk Weyl tensor, as we saw in equation (\ref{bgriemann}). Therefore, the corresponding expression for the
  Ricci tensor on    $\partial\mathcal{\bar
  M} $  will not contain a troublesome Weyl term like
  $E_{\mu\nu}$. The Ricci tensor on  $\partial\mathcal{\bar
  M} $, or equivalently, the brane, can be expressed as
\be \label{bgricci}
\mathcal{R}_{\mu\nu}(\gamma)=\frac{2\Lambda}{n-1}\gamma_{\mu\nu}-\left(\frac{1}{4M^{n-2}}\right)^2\left[T_{\mu\alpha}T^\alpha_\nu-\frac{T}{n-2}T_{\mu\nu}\right]
\ee
where
\be
T_{\mu\nu}=T^{(m)}_{\mu\nu}+2T^{(h)}_{\mu\nu}
\ee
In going from equation (\ref{ricci}) to equation (\ref{bgricci}), we
  have traded the bulk weyl term $E_{\mu\nu}$ for some holographic
  matter, $T^{(h)}_{\mu\nu}$, on the brane. So it seems that we always
  have two equivalent pictures: we can either think of the brane as moving
  in a non maximally symmetric bulk, or we can think of the brane as
  moving in a maximally symmetric bulk, provided we include some
  additional holographic matter on the brane. The remarkable thing is
  that this correspondence is exact, in that it works to all orders in
  $T_{\mu\nu}$ in equation (\ref{bgricci}). In a delightfully simple
  way, we have seen how to  extend  exact
  braneworld holography to more general braneworld geometries.
\section{On the holographic energy-momentum}
A natural question to ask is: what do know about the holographic
matter?  Unfortunately, not a great deal. For an asymptotically AdS
bulk in 5 dimensions, we might expect it to correspond to
$\mathcal{N}=4$ super Yang-Mills with the conformal invariance
strongly broken. In general, however, all we can say is that it corresponds to
some abstract quantum field theory. We {\it do} know that the
holographic matter
satisfies conservation of energy, $D^\mu T^{(h)}_{\mu\nu}=0$. This
follows from the Codazzi equation applied in the background. In
addition, we can use the contracted Bianchi identity on the brane,
$D^\mu \mathcal{G}_{\mu\nu}(\gamma)=0$,
to show that
\be \label{bianchi}
T^\mu_\alpha D_{[\mu}S^\alpha_{\nu]}=0
\ee
where
\be
S_{\mu\nu}=T_{\mu\nu}-\frac{1}{n-2}T\gamma_{\mu\nu}, \qquad T_{\mu\nu}=T^{(m)}_{\mu\nu}+2T^{(h)}_{\mu\nu}
\ee
This suggests that the holographic matter responds to changes in the
original matter content. This is no surprise, as we would expect a
change in $T^{(m)}_{\mu\nu}$ to  cause a change in the bulk Weyl
tensor. In any case, the formula (\ref{bianchi}) might offer an avenue
towards learning more about the holographic matter.

In principle we can explicitly calculate  $T^{(h)}_{\mu\nu}$ by
inverting equation (\ref{bgricci}). This would give us the holographic
energy-momentum in terms of  $T^{(m)}_{\mu\nu}$, $\gamma_{\mu\nu}$,
and $\mc{R}_{\mu\nu}$. If we wanted to relate this to the Weyl term in
the original bulk we would simply make use of equation
(\ref{ricci}). Of course, such an inversion process is highly
non-trivial. In a highly symmetric scenario such as those studied
in~\cite{Gregory:exact}, the inversion process is relatively
simple. Otherwise, we could make use of a series expansion as we will
now illustrate with an example.

Let us consider the $n$-dimensional version of a single brane
Randall Sundrum model~\cite{RS2}. We have a negative cosmological
constant in the bulk \be \label{Lambda}
\Lambda=-\frac{1}{2}(n-1)(n-2)k^2 \ee and a finely tuned brane
tension \be \label{tension} T_{\mu\nu}^{(m)}=-4M^{n-2}(n-2)k
\gamma_\mn \ee We shall now attempt to invert equation
(\ref{bgricci}) by expanding the holographic energy-momentum tensor
as a power series in $k$ \be \label{series}
T_{\mu\nu}^{(h)}=4M^{n-2}k\sum_{N=1}^\infty \tau_{\mu\nu}^{(N)}
k^{-2N}
\ee
Inserting (\ref{Lambda}), (\ref{tension}) and
(\ref{series}) into equation (\ref{bgricci}) yields the following
\bea
\mc{R}_{\mu\nu}(\gamma )&=& (n-3)\left[\tau_{\mu\nu}^{(1)}-\frac{1}{n-3}\tau^{(1)}\ga_\mn\right] \nonumber\\
&&\quad + \sum_{N=1}^{\infty}
k^{-2N}\Bigg\{(n-3)\left[\tau_{\mu\nu}^{(N+1)}-\frac{1}{n-3}\tau^{(N+1)}\ga_\mn\right]\nonumber
\\
&&\qquad\qquad-\sum_{M=1}^\infty \left[\tau_\mu^{(M)
\al}\tau_{\al\nu}^{(N+1-M)}-\frac{1}{n-2}\tau^{(M)}\tau^{(N+1-M)}_\mn\right]\Bigg\}
\eea Equating coefficients of powers of $k$, we find that
\be
\label{tau1} \tau_\mn^{(1)}=\frac{1}{n-3}G_\mn(\ga) \ee and for $N
\geq 1$, we get the recurrence relation \bea \label{taun}
\tau_{\mu\nu}^{(N+1)}&=&\frac{1}{n-3}\sum_{M=1}^\infty
\Bigg[\tau_\mu^{(M) \al}\tau_{\al\nu}^{(N+1-M)}-\frac{1}{2}\tau_\beta^{(M) \al}\tau_{\al}^{(N+1-M)\beta}\ga_\mn \nonumber\\
&&\qquad
-\frac{1}{n-2}\tau^{(M)}\tau^{(N+1-M)}_\mn+\frac{1}{2(n-2)}\tau^{(M)}\tau^{(N+1-M)}\ga_\mn\Bigg]
\eea Using (\ref{tau1}), and the recurrence relation (\ref{taun}),
we can calculate the holographic energy-momentum tensor to whatever
order we desire. We present the result here to second order \bea
\label{Th}
T_\mn^{(h)}&=&\frac{2M^{n-2}k}{n-3}\Bigg\{G_\mn(\ga)+\frac{1}{(n-2)^2k^2}\Bigg[\mc{R}_{\mu\al}\mc{R}_\nu^{\al}
-\frac{1}{2}\mc{R}_{\al\beta}\mc{R}^{\al\beta}\ga_\mn \nonumber\\
&&\qquad\qquad
-\frac{n-1}{2(n-2)}\mc{R}\mc{R}_\mn+\frac{n+1}{8(n-2)}\mc{R}^2\ga_\mn\Bigg]\Bigg\}+\mc{O}(k^{-3})
\eea If the holographic matter corresponded to a conformal field
theory, we would expect the trace of the energy-momentum to vanish.
However, in this case, the CFT is broken because the brane does not
lie on the boundary of AdS.  Taking the trace of equation (\ref{Th})
gives \be \label{trace}
T^{(h)}=M^{n-2}k\Bigg\{-\mc{R}-\frac{1}{(n-2)^2k^2}\left[\mc{R}_{\al\beta}\mc{R}^{\al\beta}
-\frac{n-1}{4(n-2)}\mc{R}^2\right]\Bigg\}+\mc{O}(k^{-3}) \ee
Although the expression (\ref{Th}) enables $T_\mn^{(h)}$ to be
determined locally on the brane, we can use (\ref{ricci}) to
substitute $\mc{R}_\mn=-E_\mn$. $E_\mn$ is  really a non-local
quantity determined by the bulk equations of motion. The
aforementioned substitution will  therefore give us a non-local
expression for $T_\mn^{(h)}$, as we might have expected. In
addition, since $E_\mn$ is traceless, we have $\mc{R}=0$. This means
that the order $k^{-1}$ term in equation (\ref{trace}) corresponds
to the trace anomaly for the (slightly broken) CFT~\cite{Shiromizu}.

For $n=5$, the bulk equations of motion have been solved order by
order to derive the solution for $E_\mn$~\cite{Kanno, deHaro}. This
solution is actually made up of a combination of  both local and
non-local pieces. In~\cite{Kanno, deHaro}, the explicitly non-local
piece is taken to be the holographic energy momentum. In contrast,
we claim that the holographic energy momentum should be given by the
BY stress-energy tensor, for the reasons discussed in the previous
section.
\section{Discussion}
In this paper we have shown how exact holography can be extended to a
general class of braneworld geometries. On the one hand we can think
of a  brane moving in a non-maximally symmetric bulk, whereas on the
other hand we can think of a brane moving in a maximally symmetric
bulk, but with some additional holographic matter on the brane. The
correspondence is exact. We can always trade a non-trivial bulk
geometry for some holographic fields on the brane. In this way, we can
always avoid the troublesome Weyl term in equation (\ref{ricci}). This
might turn out to be the most useful aspect of this holographic picture

An interesting consequence of the arguments used in this paper is that
they can trivially be extended to other gravity theories, such as
Lovelock gravity~\cite{Lovelock}. In each case, if we make use of a generalised Brown and York stress-energy tensor,
a holographic description should hold for co-dimension one
branes. In~\cite{Padilla:GB}, we studied cosmological branes moving in
a background of Gauss-Bonnet black holes. One of our conclusions was that there was
no version of {\it exact} braneworld holography, although an
approximate version did exist. I now believe this conclusion may have
been wrong. This is because we made use of the Gauss-Bonnet Hamiltonian~\cite{Padilla:ham} to
evaluate the  energy density in the bulk according to an observer on
the brane.

Let us discuss this a little further. Consider Gauss-Bonnet
 gravity described by an action $S$, including all the appropriate boundary terms~\cite{Myers}. Now suppose we wish to calculate
the quasi-local gravitational energy measured on the boundary $\partial \mc{M}$ of some spacetime region, $\mc{M}$.
We can either use the Hamiltonian, or a generalised BY stress-energy tensor. The latter is
given by the variation of the action with respect to the boundary metric
\be
T_\mn^\textrm{BY}=-\frac{2}{\sqrt{-\ga}}\frac{\partial S}{\partial \ga^\mn}
\ee
Given a Killing vector, $\zeta^\mu$, on $\partial \mc{M}$, we can still find an associated the
conserved charge given by equation (\ref{Q}). As with Einstein gravity, I believe this is a compelling reason to
adopt the Brown and York approach. Motivated by (\ref{Q}), we follow~\cite{BY}, and suggest the following formula
for the energy associated with time $t$.
\be
E=\int_{S_t} d^{n-2}\xi \sqrt{\lambda} ~u^\mu T_\mn^\textrm{BY} t^\nu
\ee
where $S_t$ are surfaces of constant $t$ in $\partial \mc{M}$, and $t^\mu\frac{\partial}{\partial y^\mu}=\frac{\partial}{\partial t}$.
Note that if we split $t^\mu$ into its lapse function and shift vector
\be
t^\mu=Nu^\mu+N^\mu
\ee
it can be shown that
\be
E=-\int_{S_t} d^{n-2}\xi N \frac{\partial S}{\partial N}+ N^\mu\frac{\partial S}{\partial N^\mu} =\int_{S_t} d^{n-2}\xi N \frac{\partial H}{\partial N}+ N^\mu\frac{\partial H}{\partial N^\mu}
\ee
where $H$ is the Hamiltonian (see~\cite{BY, Szabados} for details of the Einstein gravity case). Now, in Einstein gravity, the Hamiltonian evaluated on a solution is given by
\be \label{H}
H=\int_{S_t} d^{n-2}\xi N \frac{\partial H}{\partial N}+ N^\mu\frac{\partial H}{\partial N^\mu}
\ee
In other words, $H$ is linear in $N$ and $N^\mu$. This means that the BY approach, and the Hamiltonian approach agree on the value of the energy. However,
equation (\ref{H}) does {\it not}  hold for Gauss-Bonnet gravity. This is because, for Gauss-Bonnet gravity, the surface terms in
$H$ depend on the extrinsic curvature of $S_t$ in $\partial \mc{M}$,
 and as a result, are non-linear in $N$ \cite{Padilla:ham}.

We ought to stress, however, that even in Gauss-Bonnet gravity, the BY approach and the Hamiltonian approach agree
on the mass of black holes. This is due to the presence of a timelike Killing vector. When $t^\mu$ is Killing,  we
choose $S_t$ so that $t^\mu=Nu^\mu$, and the extrinsic curvature of $S_t$ in $\partial \mc{M}$ vanishes. This eliminates the source
of any disagreement between the two approaches, thereby explaining why they both give the same value for the black hole mass. In
contrast, a dynamical brane is generically moving around, and there
will be no timelike Killing vector. This means the (non-conserved) BY
energy will differ from the (non-conserved) Hamiltonian energy.

Given the fact that even in Gauss-Bonnet gravity, we are still able to define a conserved charge from the BY stress-energy tensor and a Killing
vector (time-like or space-like), we believe that the Brown and York approach is more reliable, although the disagreement certainly
deserves further investigation.

\vskip .5in
\centerline{\bf Acknowledgements}
I would like to thank Bernard Kay and David Wands for encouraging me
to publish this work.
\medskip
\bibliographystyle{utphys}

\bibliography{holography}

\end{document}